\documentclass[12pt,preprint]{aastex}
\usepackage{url}

\shorttitle{Mechanism of spontaneous formation of magnetic structures}

\title{Mechanism of spontaneous formation of stable magnetic structures on the Sun}

\author{I. N. Kitiashvili$^{1,2}$, A. G. Kosovichev$^{1}$ , A. A. Wray$^3$, N. N. Mansour$^3$}
\affil{$^1$Hansen Experimental Physics Laboratory, Stanford University, Stanford, CA 94305, USA \\
$^2$NORDITA, Dept. of Astronomy, AlbaNova Univ. Center, SE 10691 Stockholm, Sweden\\
$^3$NASA Ames Research Center, Moffett Field, Mountain View, CA 94040, USA}
\altaffiltext{1}{e-mail: irinasun@stanford.edu}

\begin{document}

\begin{abstract}
One of the puzzling features of solar magnetism is formation of long-living compact
magnetic structures; such as sunspots and pores, in the highly turbulent upper layer of the solar
convective zone. We use realistic  radiative 3D MHD simulations to investigate the interaction between
magnetic field and turbulent convection. In the simulations, a weak vertical uniform magnetic field
is imposed in a region of fully developed granular convection; and the total magnetic flux through the
top and bottom boundaries is kept constant. The simulation results reveal a process of  spontaneous
formation of stable magnetic structures, which may be a key to understanding of the magnetic
self-organization on the Sun and formation of pores and sunspots. This process consists of two basic steps:
1) formation of small-scale filamentary magnetic structures associated with concentrations of vorticity
and whirlpool-type motions, and 2) merging of these structures due to the vortex attraction, caused
by converging downdrafts around magnetic concentration below the surface. In the resulting large-scale
structure maintained by the converging plasma motions, the magnetic field strength reaches $\sim 1.5$~kG
at the surface and $\sim6$~kG in the interior; and the surface structure resembles solar pores.
The magnetic structure  remains stable for the whole simulation run of several hours with no sign of decay.
\end{abstract}

\section{Introduction}
Sunspots and pores represent one of the oldest and most intriguing problem of solar magnetism.
Despite the long history of observational and theoretical investigations
\citep[e.g.][]{bray1964,moore1985,brandenburg2010} the mechanism of their formation is still open.
Our understanding of MHD processes on the Sun is getting significantly improved with the rapid progress
in observational instruments, data analysis, methods and numerical modeling. For example, the data
obtained by helioseismology have provided initial information about the structure and dynamics
of convective flows around sunspots and emerging magnetic flux beneath the solar surface
\citep[e.g.][]{kosovichev2000,zhao2001,kosovichev09}. The high-resolution observations from ground-based
telescopes and Hinode space mission have provided detailed data about the filamentary magnetic structures and
flow dynamics on the surface \citep[e.g.][]{ichimoto07,bonet08,attie2009,balmaceda2010}. In addition,
important support for the understanding and interpretation of the observations is given by "realistic"
radiative MHD numerical simulations, which are based on the first
principles and take into account all essential physical processes. The recent progress
in the numerical modeling has made it possible to reproduce in simulations many observational effects
in the quiet Sun region, sunspots and active regions \citep[e.g.][]{stein2001,shussler2006,jacoutot08a,jacoutot08b,martinez08,kiti09a,kiti10},
magnetic flux emerging \citep[e.g.][]{shibata1989,cheung08,stein09} and even the whole  magnetic
structures, such as pores and sunspots \citep[e.g.][]{knolker1988,stein03,bercik2003,rempel2009}.
However, most of the modeling has been done by setting up the initial conditions with already existing
magnetic structures, e.g. a horizontal flux tube for the modeling of magnetic flux emerging, or
a vertical flux tube with strong field for the sunspot/pore structures simulations. It seems that
so far only one study succeeded in reproducing a spontaneous formation of a micropore-like magnetic
structure from an initially uniform field in the turbulent convection of the Sun \citep{stein03}.
However, the lifetime of this structure was rather short, only "few convective turnover time scales"
\citep{bercik2003}. Similar calculations by \cite{vogler2005} for a substantially shallower
convective layer did not show the structure formation.

Here, we present new results of the realistic MHD simulations that show a process of spontaneous
formation of a stable pore-like magnetic structure from an uniform magnetic field, and discuss
the physical mechanism of the structure formation, and its dynamics and evolution.

\section{Numerical setup}

For the simulations we used a 3D radiative MHD code, "SolarBox", developed by A. Wray at NASA
Advanced Supercomputing Division \citep{jacoutot08a,jacoutot08b,kiti09b}. The code is built for
3D simulations of compressible fluid flows in a magnetized and highly stratified medium
of top layers of the convective zone and the low atmosphere, in the rectangular geometry.
The code has been thoroughly tested and  compared for test runs with a similar code of \cite{stein2001}.
The code is based on the Large-Eddy Simulation (LES) approach,
and solves the grid-cell averaged equations of the conservation of mass, momentum and energy.
It takes into account the real-gas equation of state, ionization and excitation of all abundant
species, and magnetic effects. A unique feature of the code is implementation of various
sub-grid scale turbulence models. For this particular simulations we use the minimal hyperviscosity
model \citep{jacoutot08a}.

The simulation results are obtained for the computational domain of $6.4\times6.4\times5.5$~Mm with
the grid sizes: $50\times50\times43$~km, $25\times25\times21.7$~km and $12.5\times12.5\times11$~km
($128^2\times127$, $256^2\times253$ and $512^2\times505$ mesh points). The domain includes a top,
5~Mm-deep, layer of the convective zone and the low atmosphere. The lateral boundary
conditions are periodic; and the top and bottom boundaries were closed for flows and maintain the constant
total magnetic flux. The results have been verified by increasing the computational domain size
to 12.8 Mm in the horizontal directions. The initial uniform magnetic field of various strength 1, 10
and 100~G, was superimposed over the fully developed granular convection. The computation runs were
up to 8~hours of solar time.

We describe first the modeling of the solar convection without magnetic field, which prepares
the initial conditions, and then, the simulations with the superimposed weak vertical magnetic field.

\subsection{Simulation of the quiet Sun regions}

\begin{figure}
\begin{center}
\includegraphics[scale=1.5]{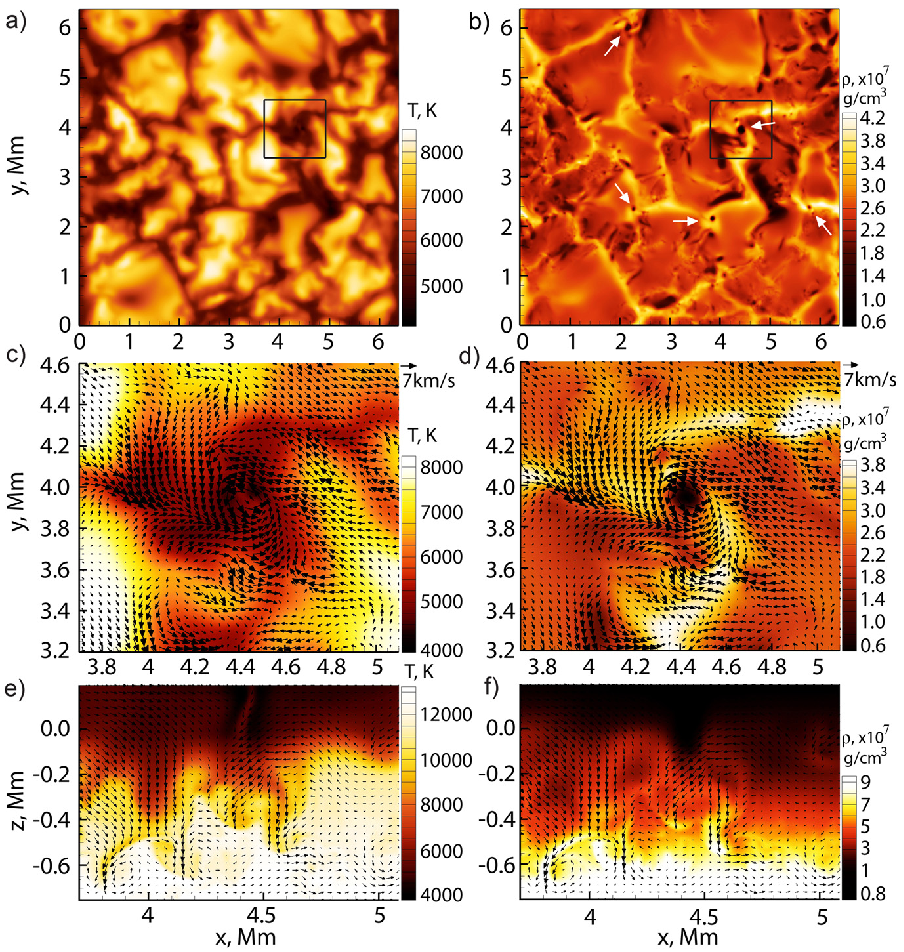}
\end{center}
\caption{Snapshots of granular convection at the surface for the simulations without magnetic field,
and the horizontal resolution of 12.5~km: temperature (left column) and density (right).
The black square indicates a large whirlpool, the horizontal and vertical structure of which are shown
in panels c) -- f). Black arrows show the flow velocity. White arrows in panel b) point to the centers
of some vortices (dark low-density points).
\label{noMHDswirls}}
\end{figure}

Figure~\ref{noMHDswirls} shows snapshots for temperature (left column) and density (right) at the surface
for the case without magnetic field. As usual, the convective motions develop a characteristic granulation
pattern with the relatively hot and less dense upflowing plasma in the middle of the granular cells, and
the lower temperature and higher density downflowing plasma at the intergranulation boundaries (dark
lines of granulation). An interesting feature of the convective flows is the formation of whirlpool-like
motions of different sizes ($\sim 0.2-1$~Mm) and lifetimes ($\sim 15 - 20$~min\footnote{
The lifetime can be longer. It is difficult to estimate it accurately because during its evolution a vortex
can significantly change the shape and size, and sometimes almost disappear and then rise again.})
at the vertexes of the intergranular lanes. The vortical motions are particularly well seen in the density
variations. The centers of the whirlpools are seen as dark dots (indicated by white arrows
in Fig.~\ref{noMHDswirls}b) in the intergranular space. The evolution of these vortices is ultimately
related to the dynamics of convective motions in the domain. The convective flows may sometime collect
swirls in a local area, then merge and destroy them. Such vortical structures in simulations were
first described by \cite{stein1998}. They showed that stronger vortices usually correlate with downflows,
and this is also found in our results.

From time to time, convection creates pretty big whirlpools, as the one indicated by square in Fig.~\ref{noMHDswirls},
which can swallow up other smaller swirls around them. The big swirls are usually easy to see also
in the surface temperature and intensity variations. The detailed structure of a large whirlpool is shown
in Figs~\ref{noMHDswirls}c-f. The whirlpool structure is characterized by: 1) formation of ''arms'' of higher
density that correlates with lower temperature; 2) a pronounced vortical structure of the velocity flow;
3) increased magnitude of the horizontal velocity up to 7 -- 9~km/s; 4) a sharply decreased density
in the central core of the vortex, and a  slightly higher temperature than in the surrounding. The typical depth
of large swirls is about 100 -- 200~km. Inside the whirlpool shown in Fig.~\ref{noMHDswirls}e, we can see
a higher temperature tube-like structure, but it is unstable (in comparison with the whirlpool lifetime),
and can be destroyed during the swirling motions.

The vortical motions in the solar granulation have been detected in high-resolution observations
\citep[e.g.][]{potzi2005}, and the observational results generally agree with the simulations. In particular,
recent observations of a quiet region, near the solar disk center detected magnetic bright points following
a logarithmic spiral trajectory around intergranular points and engulfed by a downdraft \citep{bonet08}.
The observations were interpreted as vortical flows that affects the bright point motions. These
whirlpools have the size $\lesssim 0.5$~Mm and the lifetime of about 5~min, without preferred direction
of rotation. The distribution of vortices studied from the ground \citep{wang1995,potzi2005,potzi2007,bonet08}
and space observations \citep{attie2009} shows strong preferences to concentration in regions of downflows,
particulary at the mesogranular scale \citep{potzi2005,potzi2007}. Our simulations for the domain of the
horizontal size of 12.8~Mm also show a tendency of concentration of vortices on a mesogranular scale.
We plan to discuss this effect in a separate paper. Here we focus on the links between the whirlpools
and magnetic structure formation.

\subsection{Spontaneous formation of magnetic structures}

\begin{figure}
\begin{center}
\includegraphics[scale=1.5]{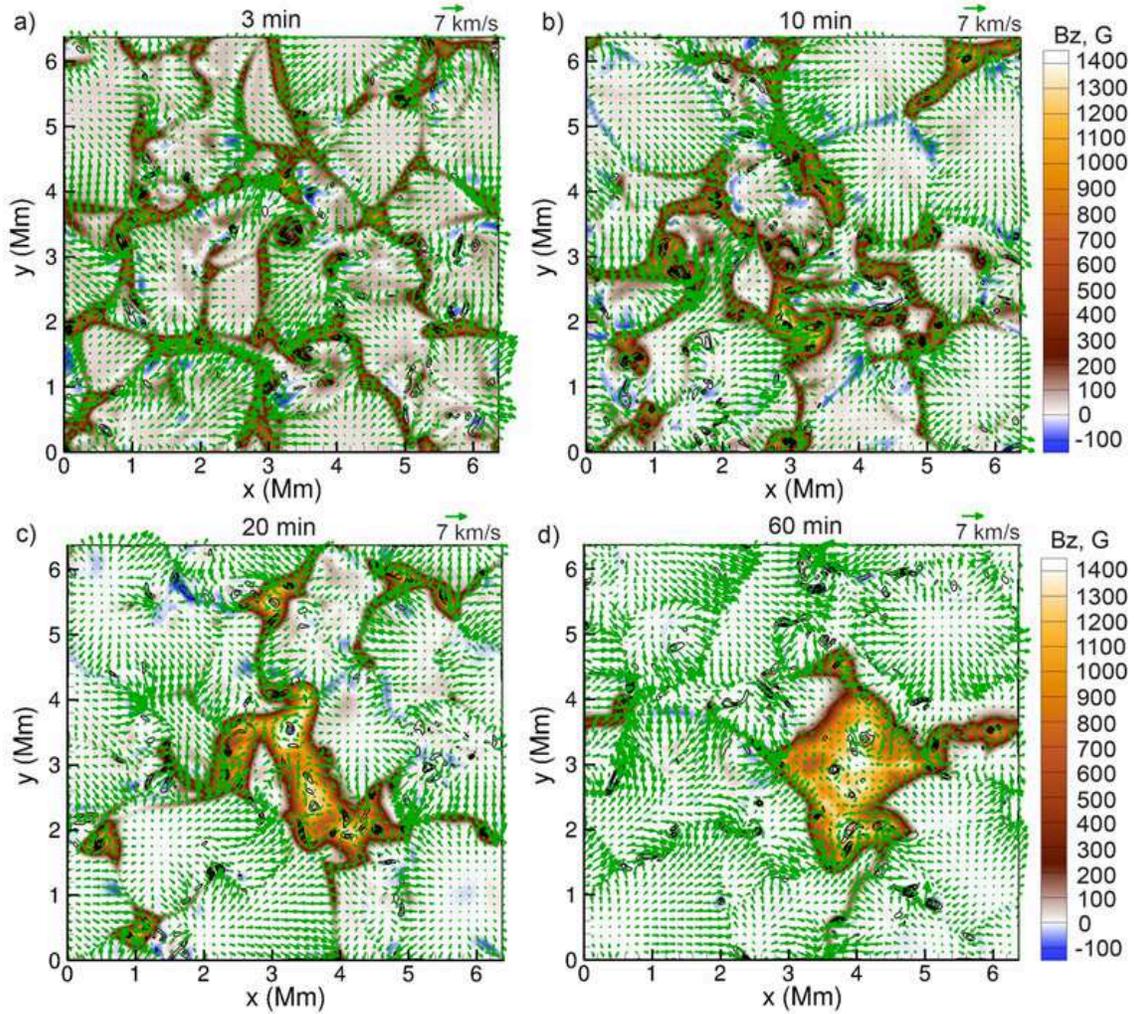}
\end{center}
\caption{Snapshots of the surface distribution of vertical magnetic field (color background), horizontal
flows  (arrows) and vorticity magnitude (black contour lines) for four moments of time: 3, 10, 20 and 60 min,
from the moment of initiation of a uniform magnetic field ($Bz_0=100$~G).
\label{B100_256ls_xy}}
\end{figure}

\begin{figure}
\begin{center}
\includegraphics[scale=0.9]{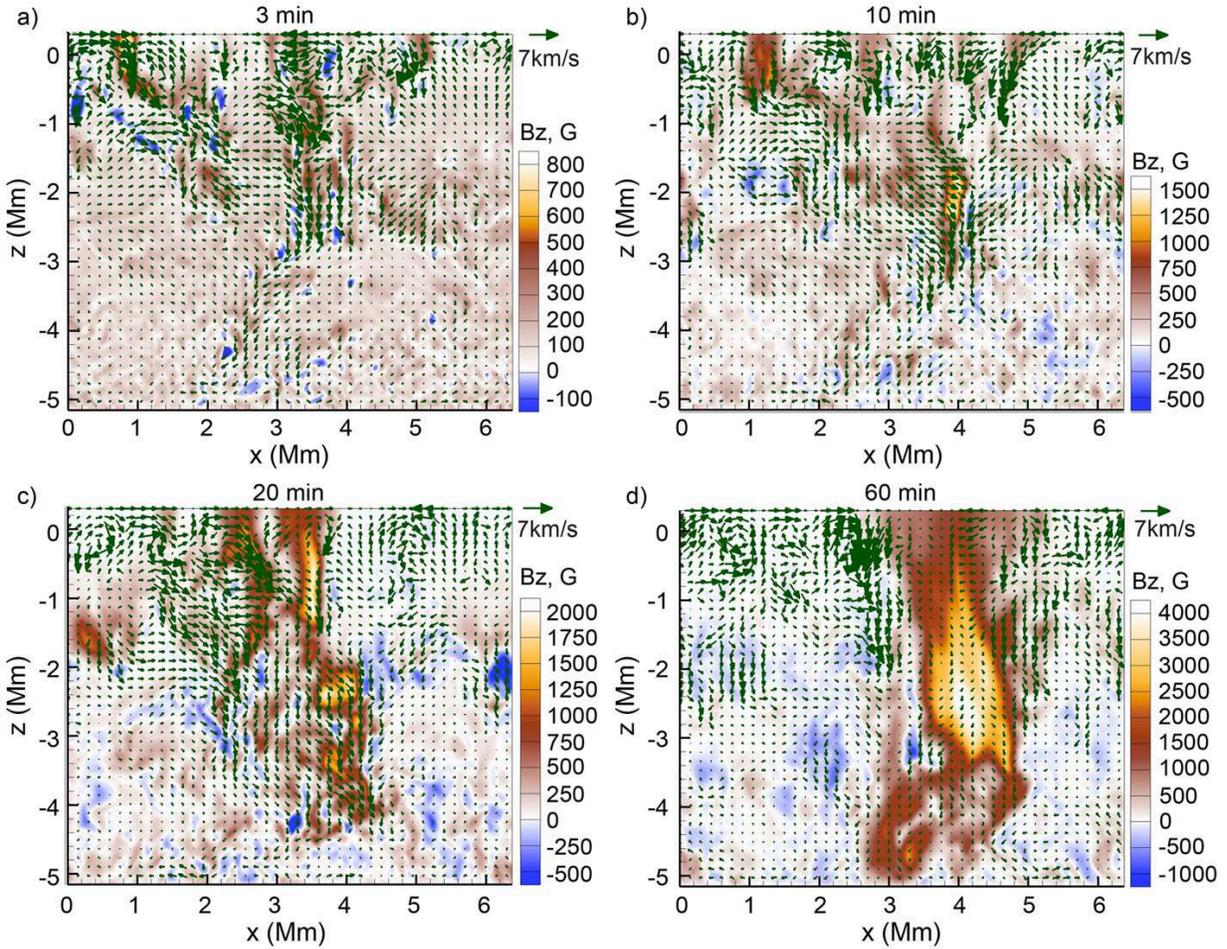}
\end{center}
\caption{Vertical snapshots of the vertical magnetic field (color background) and the horizontal
flows  (arrows) for $t=$~3, 10, 20 and 60 min.
\label{B100_256ls_xz}}
\end{figure}

To investigate the process of magnetic field structuring in the turbulent convective plasma we made
a series of simulations for the initial vertical uniform magnetic field, $Bz_0$, varying from
1 to 100~G, different computational grids and domain sizes. Qualitatively the simulation
results are very similar in all these cases, and show the formation of stable magnetic pore-like
structures. In Figures~\ref{B100_256ls_xy} and~\ref{B100_256ls_xz} we present the results for the case
of $Bz_0=100$~G, the grid size of 25~km, and the domain size of $6.4\times6.4\times5.5$~Mm,
for which we have made the longest run ($\gtrsim 8$hours). The periodic lateral boundary conditions
allow us, for the illustration purpose, to shift the horizontal frame so that the structure is located
close to the center. As we see in the simulations, the structure can be formed in any place of our
computational domain, but usually the process starts at one of the strongest vortices.

Figure~\ref{B100_256ls_xy} shows snapshots of the vertical magnetic field (background color image),
horizontal flow field (arrows), and the vorticity magnitude (black contour lines) for four moments of time:
3, 10, 20 and 60~min after the moment $t=0$, when the 100~G vertical field was uniformly distributed in
the computational domain. During few minutes the magnetic field is swept into the
intergranular lines, and is significantly amplified up to $\sim 500-1000$~G. The vortices and magnetic
field get concentrated at some locations in the intergranular lanes, where they are deformed and
became elongated (or elliptically shaped) along the intergranular lines. The process of formation of
a large-scale magnetic structure starts at a strongest vortex in our domain. Since our computational
domain is periodic we choose the horizontal coordinates in such a way that this vortex is located in
the middle of Figure~\ref{B100_256ls_xy}a. This large swirl sweeps magnetic field and also becomes stretched
by strong horizontal shear flows. The whirlwind causes deformation of the intergranular space, and creates
a cavity of low density, temperature and pressure. The cavity expands and increases the accumulation of magnetic
field (Fig.~\ref{B100_256ls_xy}a). A similar process of magnetic field concentration, sweeping, twisting and
stretching by vortical motions in the intergranular lane was initially observed in the simulations of \cite{stein03}.

During the next few minutes the deformation of the "parent" vortex continues; then it gets destroyed
on the surface by $t=10$~min (Fig.~\ref{B100_256ls_xy}b), but leaves strong downdraft motions in the interior
(Fig.~\ref{B100_256ls_xz}b). The process of accumulation of magnetic flux in this area continues.
The local concentrations of magnetic field and vorticity get stronger and are moved by convective
motions in the direction of the initial cavity, into the region where the gas pressure remains systematically
low due to the downdrafts. In Fig.~\ref{B100_256ls_xy}b several small-scale structures can be seen in this area.
However, it is difficult to recognize the center of the attraction on the surface. Collisions of flows coming
from different directions create additional vortices, and this seems to accelerate the accumulation process.
As a result, the different small magnetic structures join together in a magnetic conglomerate that continues
to attract other magnetic micro structures (Fig.~\ref{B100_256ls_xy}c) and becomes more compact
(Fig.~\ref{B100_256ls_xy}d). This process of the magnetic structure formation is particularly well illustrated
in the movie\footnote{See also movie: \url{http://soi.stanford.edu/~irina/fig2swirls_movie.mpeg}}.
It shows a correlation between the distribution of vortices and the areas with concentrated
magnetic field elements, as well as the mutual influence of convection, vortices and magnetic field.

\begin{figure}
\begin{center}
\includegraphics[scale=0.9]{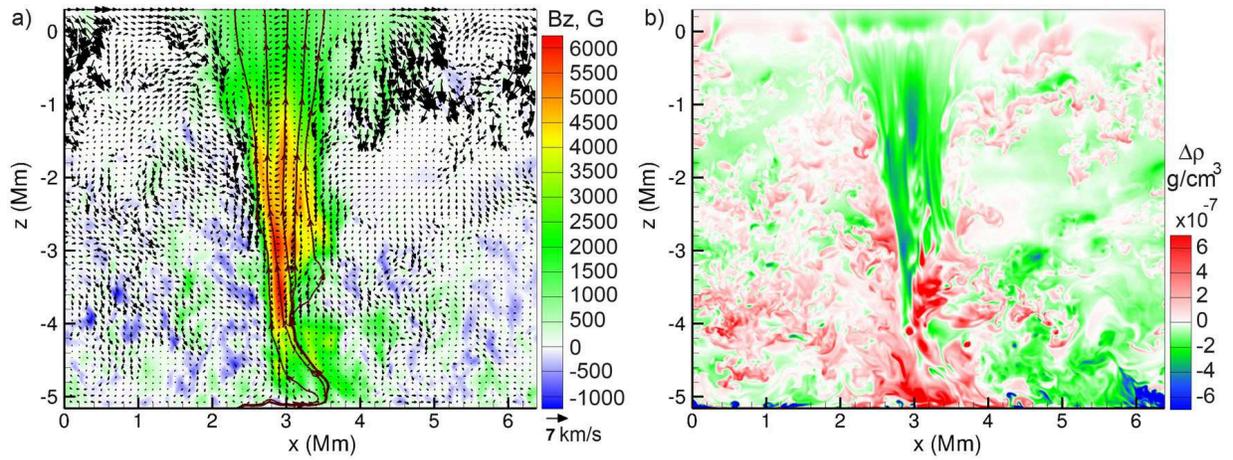}
\end{center}
\caption{Vertical slices though the magnetic structure at t= 84 min for the 12.5~km resolution:
a) the vertical component of magnetic field (color background), flow velocity (arrows),
magnetic field lines (contour lines); and b) variations of density with respect to a mean density profile
of the convection simulations without magnetic field. The velocity vectors and magnetic field lines
in panel a) are calculated from the corresponding $x$- and $z$-components.
\label{B100_512ls_xz}}
\end{figure}

Figure~\ref{B100_256ls_xz} shows a vertical slice for the same moments of time. The arrows show the
velocity field calculated from the $x$- and $z$-components. Thus, we can see that the initially
uniform magnetic field has local concentrations in the regions of the near-surface downdraft, and has a very
fragmentary structure during the first few minutes (Fig.~\ref{B100_256ls_xz}a). The continuing and
extending into the deeper layers downflows are accompanied by the local concentrations of the magnetic field
strength (Fig.~\ref{B100_256ls_xz}b), which grows after $\sim20$~minutes to $\sim2$~kG below the surface
(Fig.~\ref{B100_256ls_xz}c). By $t=60$~min, the magnetic field is mostly concentrated in a single
flux structure with a maximum field strength of about $\sim 4$~kG at the depth of $\sim 1-4$~Mm, and $\sim1.4-1.5$~kG
at the surface (Fig.~\ref{B100_256ls_xz}d). The magnetic field is weaker and more disperse near the bottom
of our domain, which is impenetrable for flows. The flux-tube interior represents a cluster-type structure
(Figs.~\ref{B100_256ls_xy}d,~\ref{B100_256ls_xz}d and~\ref{B100_512ls_xz}), as initially predicted by
\cite{parker1979} and observed on a larger scale for a sunspot by helioseismology \citep{zhao2010}. In our simulations
the cluster structure is represented by internal field concentrations (flux tubes), $100-200$~km thick,
in which the field strength reaches 6~kG after 1 hour (Fig.~\ref{B100_512ls_xz}a). The velocity distribution
shows strong, often supersonic, downflows around the magnetic structure. Inside the magnetic structure
the convective flows are suppressed by strong magnetic field. However, despite the weak velocities
($\sim0.1-0.2$~km/s) there are very small elongated convective cells resembling the umbral dots observed
at the surface. The distribution of density fluctuations (Fig.~\ref{B100_512ls_xz}b) shows the following basic
properties: a) decrease of density inside the magnetic structure; b) fine needle-like structurization; c) a thin
near-surface layer of slightly higher density; and d) higher density around the structure, particularly in the
deep layers of the domain  (Fig.~\ref{B100_512ls_xz}b). We have followed the evolution
of the magnetic pore-like structure for more than 8 solar hours, and did not see any indication
of its decay. However, the shape and other properties fluctuate during this evolution. The decay of this
structure is probably prevented by keeping the total magnetic flux constant during the run.
We repeated the simulation when the initial uniform magnetic field was introduced at the different moment
of time. In one case we observed the formation of two separate magnetic structures, which later merged together,
but the whole process was very similar.

\section{Conclusions}

Initially, the vortical motions in quiet-Sun regions were detected on a large scale of $\sim 5$~Mm \citep{brandt1988}.
With the development of instrumentation it became possible to observe small-scale swirls in the photosphere
\citep{bonet08} (with size $< 0.5$~Mm) and also in the chromosphere $\sim 1.5$~Mm \citep{wedemeyer09}.
Very recently, the high-resolution observations revealed a process of dragging of small-scale magnetic
concentrations toward the center of a convective vortex motion in the photosphere \citep{balmaceda2010}.

Our simulations show that the small-scale vortices representing whirlpool-type motions at intersections
of the intergranular lanes may play important roles in the dynamics of the quiet-Sun and magnetic regions.
Our simulations show that the process of formation of small-scale magnetic structures and their accumulation
into a large-scale magnetic structure is associated with strong vortical downdrafts developed around these structures.
The resulting stable pore-like magnetic structure has the highest field strength of $\sim6$~kG at the depth
of $1-4$~Mm and $\sim1.5$~kG at the surface. It has a cluster-like internal structurization, and seems to be maintained
by strong downdrafts converging around this structure and extending into the deep layers. Our simulations
show that this internal dynamics plays a critical role in the magnetic self-organization of solar magnetic
fields and formation of large-scale magnetic structures.


\end{document}